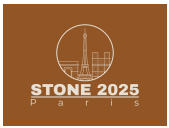

# COLISEUM PROJECT: CORRELATING CLIMATE CHANGE DATA WITH THE BEHAVIOR OF HERITAGE MATERIALS

A. Cormier[1*] and D. Roqui[2], F. Surma[3], M. Labouré[4], J. Vallet[5] & O. Guillon[6], N. Grozavu[7], A. Bourgès[8]

**Abstract**

Heritage materials are already affected by climate change, and increasing climatic variations reduces the lifespan of monuments. As weathering depends on many factors, it is also difficult to link its progression to climatic changes. To predict weathering, it is essential to gather climatic data while simultaneously monitoring the progression of deterioration. The multimodal nature of collected data (images, text…) makes correlations difficult, particularly on different time scales. To address this issue, the COLISEUM project proposes a methodology for collecting data in three French sites to predict heritage material behaviour using artificial intelligence computer models. Over time, prediction models will allow the prediction of future material behaviours using known data from different climate change scenarios by the IPCC (Intergovernmental Panel on Climate Change). Thus, a climate monitoring methodology has been set up in three cultural sites in France: Notre-Dame cathedral in Strasbourg (67), Bibracte archaeological site (71), and the Saint-Pierre chapel in Villefranche-sur-Mer (06). Each site has a different climate and specific materials. In situ, microclimatic sensors continuously record variations parameters over time. The state of alteration is monitored at regular intervals by means of chemical analyses, cartographic measurements and scientific imaging campaigns. To implement weathering models, data is gathered in alteration matrix by mean of a calculated weathering index. This article presents the instrumentation methodology, the initial diagnostic and the first results with the example of Strasbourg Cathedral site.

**Keywords**: Cultural heritage materials, Climate change, Damage assessment, Artificial intelligence, Data collection

## 1. Introduction

The intensification of climatic variations reduces the lifespan of monuments (Burke et al., 2011). This issue, raised since 1996 (Rowland, 1996), was particularly highlighted at the UNESCO World Heritage Convention in 2007. In 2020, the 20e General Assembly of ICOMOS (International Council on Monuments and Sites) declared a climatic and ecological urgency (Rao, 2020). Since the 2000s, studies on the impact of climate change on heritage have already been conducted, demonstrating the need to quantify current and future changes in the behavior of heritage materials. The study presented in this article is part of the COLISEUM project, which aims to assess the deterioration of materials in built heritage under the influence of climatic and environmental conditions. This approach is based on a multi-disciplinary methodology integrating in situ instrumental measurements, laboratory analyses and numerical modelling. The main aim of this initial study is to establish a climate and damage monitoring methodology. This article presents the case study of Strasbourg cathedral site: climatic instrumentation, initial damage assessment and monitoring via the conception of a weathering matrix.

### 1.1. Notre-Dame of Strasbourg Cathedral site and materials

Strasbourg Cathedral, emblematic gothic monument, was built between 1015 and 1439 with Vosges sandstones. Its classification as a Historic monument in 1862 and its inclusion on UNESCO's World Heritage List in 1988 witnesses its architectural importance. This study focuses on the cathedrals spire (142 m) octagonal base at a height of 100 metres. Different Vosges sandstones were used in the construction of the cathedral, each with different petrophysical properties (Thomachot, 2002). Sandstone is a detrital sedimentary rock, formed by the aggregation of predominantly sandy grains (0.063mm to 2mm) and consolidated during diagenesis. Weathering of sandstone, due to both natural and man-made factors, is a major problem affecting Strasbourg Cathedral (Colas, 2011), especially in the last 42 metre of the spire exposed to extreme weather conditions. Climatic conditions, such as freeze-thaw cycles, precipitation and atmospheric pollution, all contribute to the gradual degradation of sandstone. This deterioration manifests itself in various sandstone pathologies, which can be classified according to the ICOMOS glossary (2008) into five sub-categories: cracks and deformations, detachment, loss of material, chromatic alteration and deposits, biological colonization.

[1*] A. Cormier, PhD student, ETIS, C2RMF, CICRP, EPITOPOS, adele.cormier@epitopos.fr

[2] D. Roqui, PhD student, C2RMF, ETIS, david.roqui@ensea.fr

[3] F. Surma, conservation scientist & CEO EPITOPOS, fabrice.surma@epitopos.fr

[4] M. Labouré, heritage restorer & CEO Mescla Patrimoine, m.laboure@mescla.eu

[5] J. Vallet, conservation scientist, CICRP, jean-marc.vallet@cicrp.fr & O. Guillon [6], photographer, odile.Guillon@cicrp.fr

[7] N. Grozavu, full professor, ETIS-CNRS UMR 8051, ENSEA, CY Paris Université, nistor.grozavu@cyu.fr

[8] A. Bourges, conservation scientist, C2RMF, ann.bourges@culture.gouv.fr

## 1.2 Climatic conditions and heritage material alteration in Strasbourg

Strasbourg's climatic environment is characterized by semi-continental conditions, with well-defined seasons and precipitation distributed throughout the year. It has wide annual and daily temperature ranges, regular rainfall and prevailing north/north-easterly winds. The main parameters of Strasbourg's climate are temperatures between 0 and 5°C in winter, rising to over 30°C in summer. Relative humidity fluctuates between 65% in July and 84% in November, with saturation peaks. There is about 964mm of precipitation per year with frequent thunderstorms in summer. The studied microclimate of the cathedral spire is even more extreme, given the location of the climate sensors on the cathedral spire at a height of 100m. Climate change is defined by the IPCC (Intergovernmental Panel on Climate Change) as changes in the mean or the variability of its properties for an extended period. It is already visible in the Bas-Rhin region. There is a significant reduction in the number of days of frost and snow per year. These variations in climatic cycles modify material weathering kinetics. Therefore, this project proposes a methodology for systematically collecting data to monitor weathering. Data will be used in artificial intelligent *Transformer* models (Vaswani et al., 2017) to predict the behaviour of heritage materials, as shown in figure1. The aim is to determine how to link together collected multimodal data (numbers, pictures, text) at the scale of heritage material to correlate climate change and weathering.

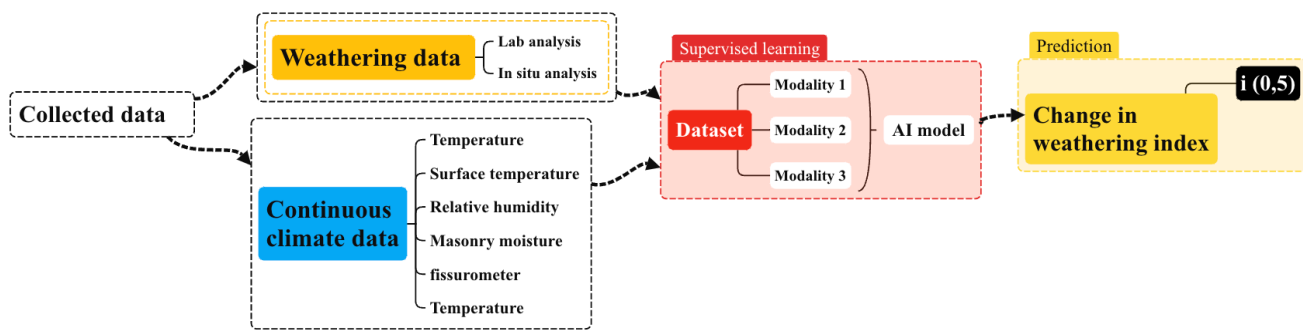

*Figure 1: Schematic summary of the monitoring and data collected on the spire of Strasbourg's Notre-Dame cathedral since April 2024, for use in predictive algorithms of alteration*

## 2. General methodology

### 2.1 Step 1: Selection of areas and instrumentation for continuous climate monitoring

This methodology consists in non-destructive and simultaneous monitoring of both climate and the state of alteration of studied site. The instrumentation and the first on-site campaign presented were carried out in April 2024. As represented on figure 2, two contrasting zones of the cathedral spire were selected. The Northeast face less altered and less exposed to direct sunlight and the South-west face more altered and subject to strong thermal variations. Blocks of control batches were also placed on the spire (and on all the study sites) for natural ageing over 3 years. The batches are 5x5x5cm cubes of granite (Bibracte), Alsatian sandstones (Vosges sandstone, Bitburg sandstone, Staub sandstone), Barutel limestones (Sèle quarry, Nîmes), Estaillade limestones (Provence quarries, Oppède). For continuous climate monitoring, sensors have been installed for humidity (%) and ambient temperature (°C), surface temperature (°C) for dew point monitoring and relative humidity in the masonry (water circulation monitoring). To interpret differences in weathering, sensors are positioned in each zone, and in different orientations. The location of the sensors is shown in figure 2 and specified in table 1.

*Table 1: Type and measurements of sensors installed on the two exposures of the cathedral spire*

| Sensor | Measure | Resolution | Measurement frequency |
|---|---|---|---|
| Thermo-hygrometer | Temperature °C | ± 0,1 °C | 20 minutes |
|  | Relative humidity % | ± 1,5 % |  |
| Surface probe PT100 | Surface temperature °C | ± 0,1 °C | 20 minutes |
| TDR Moisture probe | Water content (10cm depth) | ± 1,0 % | SD card storage |
| Fissurometer | Crack size | ± 0,1mm | 60 minutes |
|  | Temperature °C | ± 0,1 °C |  |
|  | Relative humidity % | ± 1,0 % |  |

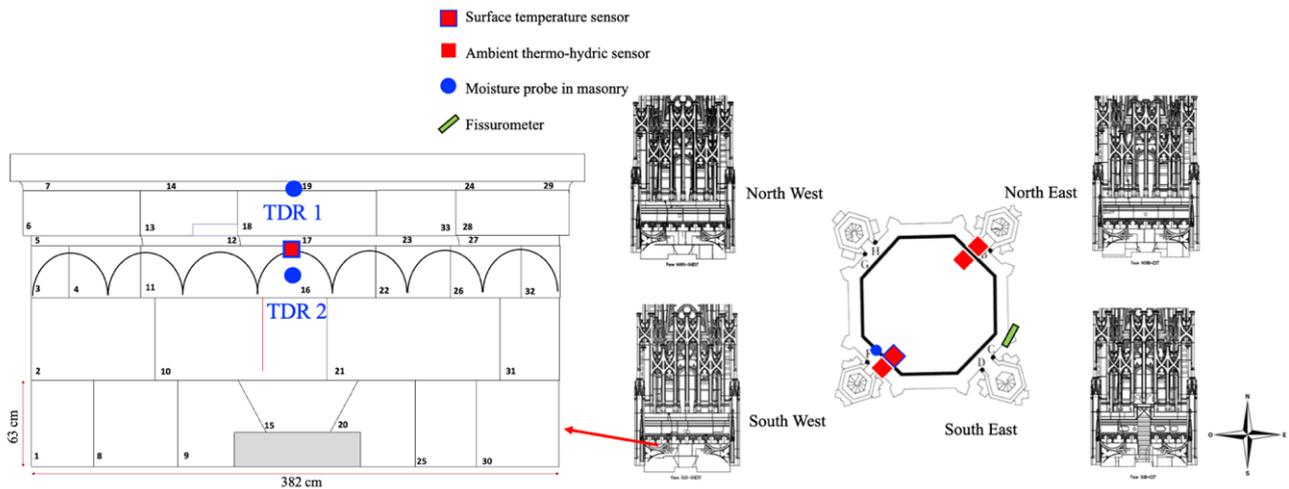

*Figure 2: Location of sensors on the cathedral spire installed in April 2024*

## 2.2 Step 2: Punctual collection of weathering data

To correlate climatic data with the evolution of weathering, an initial damage assessment is defined to start. This initial damage assessment is based on four types of point-by-point diagnosis, supplemented by laboratory and structural analyses, pictured in figure 3. Structural parameters include the consideration of geometry, location and type of material. In situ, surface analysis campaigns involve the use of portable, non-destructive analysis techniques such as colorimetry, relative surface humidity (capacitive moisture meter) and imaging campaigns. Imagery is used to visually monitor each area. An alteration index between 0 and 5 is assigned according to ICOMOS categories of stone deterioration patterns. This campaign is supplemented by laboratory analysis of collected samples: analysis of soluble salts with ion chromatography, gravimetric measurement of water content in masonry and analysis of efflorescence by X-ray diffraction (XRD). Hygroscopic water content, which corresponds to the maximal quantity of water adsorbed by the sandstone, is also measured. This value corresponds to the maximal quantity of water adsorbed by the sandstone and gives an indication on the presence of soluble salts. Finally, this initial campaign is repeated every 3 to 6 months. Collected data is gathered in a grid of analysis points (one per sandstone block). Subdividing each face into sub-zones enables the compilation of results in an alteration matrix that will feed the predictive models.

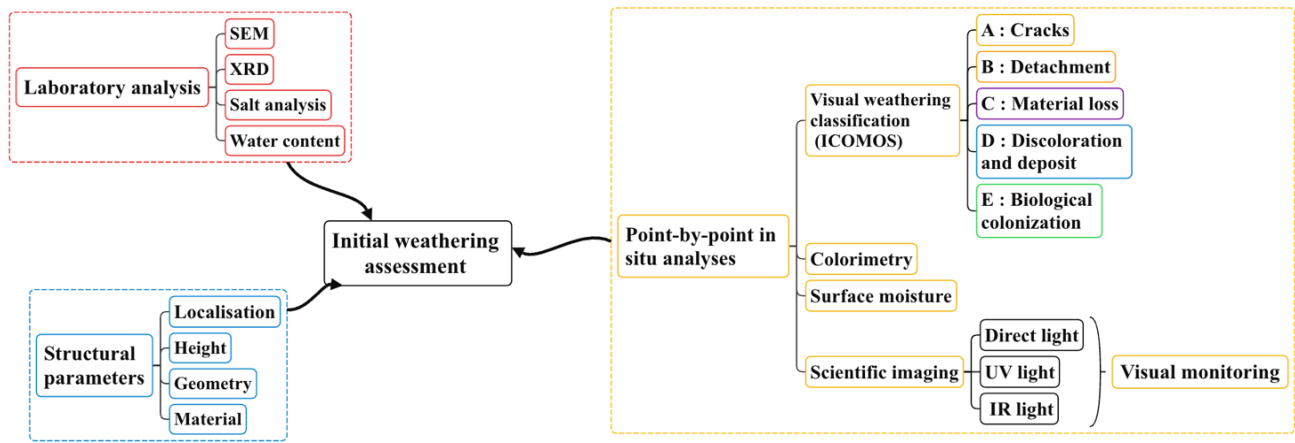

*Figure 3: Parameters and analyses included in the initial damage assessment*

## 2.3 Step 3: Setting up local weathering index and matrix

Various methodologies have already been proposed for damage classification systems and degradation criteria to quantify the visual deterioration of cultural heritage (Bonazza & Sardella, 2023). These classifications consist of categorizing defects according to a scale of discrete variables (Damas et al, 2022), from the most favourable condition to the most unfavourable (generalized deterioration). For example, a vulnerability index has been set up by (Ortiz&Ortiz 2016) using the alterations from the ICOMOS glossary. Regarding COLISEUM project, as the predictive computer model is fed by a matrix of parameters, the calculation of a weathering index as a continuous variable is compulsory.

## 3. First results: data processing and alteration matrix conception

### 3.1 Processing continuous climate data

Initial climate monitoring between April and September 2024 already reveals a particularly extreme microclimate on the Strasbourg Cathedral's spire, with significant daily variations in temperature and humidity. A maximum temperature rise of 39,6°C is reached. Results presented on table 2 show a very high temperature differential, especially for a spring/summer season. The same applies to relative humidity. Between May and September 2024, it exceeded 90% on 94 out of 118 days. Monitoring the surface temperature between April and August 2024 follow those of the air with a certain delay, which influences the expansion/contraction cycles of the materials. This surface temperature differential with high humidity causes the dew point temperature to be regularly exceeded (more than 200 times in 4 months). This gives rise to condensation/evaporation cycles, leading to the formation of surface and deep-seated disorders, mobilizing water in the masonry. A close eye must be kept on these cycles, as well as on the evolution of salts. The microclimates of the two faces are different, with higher measured temperatures on the Southwest side (average temperature is one degree higher on the South West than that of the North East face over the period).

*Table 2: Climatic data recorded between 17/05/2024 and 30/08/2024*

| Average T °C | Average Ts °C (Surface) | Average HR % | Tmax °C | Tmin °C | HRmax % | Hmin % |
|---|---|---|---|---|---|---|
| 19,8 | 20,9 | 70,5 | 39,6 | 9,2 | 100,0 | 29,5 |

Tracking a crack in the gallery at the base of the spire reveals its progressive enlargement (graph (a) figure 4). The evolution of the crack will be correlated with specific climatic events through time. Concerning masonry, according to graph (b), the top probe TDR1 measures a higher water content than the bottom probe TDR2. Given the structure of the studied faces overhung by a cornice (with a lead and plaster cap), the circulation of water through the top is expected. Regarding waterflow in masonry, whatever the face studied, sandstone moisture seems to originate from water flowing from above (base of the spire). This water may be loaded with solubilized salts, which recrystallize at depth and on the surface during the soaking/drying cycles observed. Water content in masonry decreases over the summer, particularly on the lower probe, which is more exposed to the sun. Studying the presence of soluble salts will help to understand the influence of water circulation on weathering.

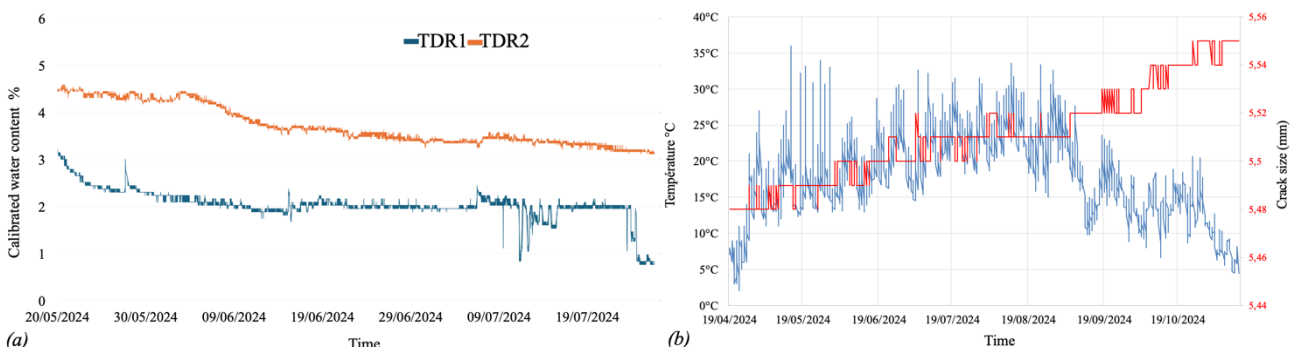

*Figure 4: (a) Evolution of water content measured at 10cm in the masonry (TDR 1 high TDR 2 low) between May and July 2024; (b) Changes in crack size and ambient relative humidity between April and October 2024*

### 3.2 Laboratory and punctual analysis processing

Analysis of first intervention in April 2024 represents the initial state of studied materials. Results show advanced conditions of deterioration on both faces, particularly Southwest. On site, spot measurements of surface humidity with a *trotec BM31* capacitance hygrometer and colorimetry with a *colorcatch nano* also reveals disparities. Surface humidity maps were produced, confirming a higher humidity at the top of each face. Analyses of the water and soluble salt content of five areas at different drilling depths identified the presence of soluble salts. Results are presented in table 3 with the example of a drilling area. It shows a high level of salt contamination (chlorides, sulphates, nitrates). Indeed, the permissible salt content of a stone is 0.1% for chlorides or sulphates (excluding gypsum) and 0.5% for nitrates. In addition, powders with a high hygroscopic water content (in bold table 3) are the most contaminated with sulphates, which confirms the presence of sulphated hygroscopic salts. Salts are mainly located at depth, as are the wettest areas ( more than 5% humidity). These observations suggest that water originates from inside the masonry, then exits through the pores and evaporates, forming efflorescence. X-ray diffraction analysis of efflorescence identified gypsum ($CaSO_4.2H_2O$), potassium salts ($KNO_3$, $K_2SO_4$) and sodium sulphates (thenardite $Na_2SO_4$). Particularly high levels of humidity (up to 9%) were found on the south-west face. This shows a very advanced initial state of weathering which will be monitored from there.

*Table 3: Results of the determination of soluble anion and cation salts, of water content (w) and of hygroscopic water content ($w_h$) of powders taken in the middle of the South West face at three different depths (drilling S1)*

| Depth | Chloride % | Nitrate % | Sulfate % | Sodium % | Magnesium % | Calcium % | w % | $w_h$% |
|---|---|---|---|---|---|---|---|---|
| 0-1cm | 0,01 | 0,09 | **1,86** | 0,04 | 0,01 | 0,97 | 1,2 | 5,4 |
| 1-4cm | 0,02 | 0,1 | **5,51** | 0,09 | 0,02 | 2,31 | 4,7 | **10,0** |
| 4-6cm | 0,01 | 0,06 | **0,47** | 0,08 | 0 | 0,38 | **9,2** | **13,2** |

All collected data will be used to fill in the weathering matrix following a precise methodology and will eventually feed prediction models.

### 3.3 Weathering index calculation and construction of the matrix

An alteration map of the Southwest face and the general structure of associated weathering matrix is shown in figure 5. Each row of the matrix corresponds to a sub-zone, and each column to an associated weathering or climatic parameter (more than 40 parameters). Thus, the calculated index i reflects the state at which climate monitoring began and is calculated as followed. The first step is to delimit and label precisely the sub-zones to be mapped. For the Notre-Dame de Strasbourg spire, they are the 70 sandstone blocks and the 8 batches blocks (78 lines). Labelling is important for linking the modalities together in the algorithm (weathering, pictures, sensors and weathering parameters). The second step is to fill in the invariable "structural" information for each block, such as sandstone type, height or block configuration. This results in a structural index $i_{structure}$. Thirdly, for each subzone alterations are recorded and assigned a sub-index (between 0 and 5) according to the degree of affection, represented table 4. Rating from 0 to 5 depends on the percentage of coverage of the area for each sub-family of alteration. Additionally, measured parameters from in situ analyses (colorimetry, surface humidity) and laboratory analyses are added. In the end, a continuous variable $i_{alteration}$ is obtained and, at last the general weathering index *i* is calculated by averaging all these sub-indices. Over time, the matrix will be completed with climate monitoring parameters. Each block is associated with one or more sensors.

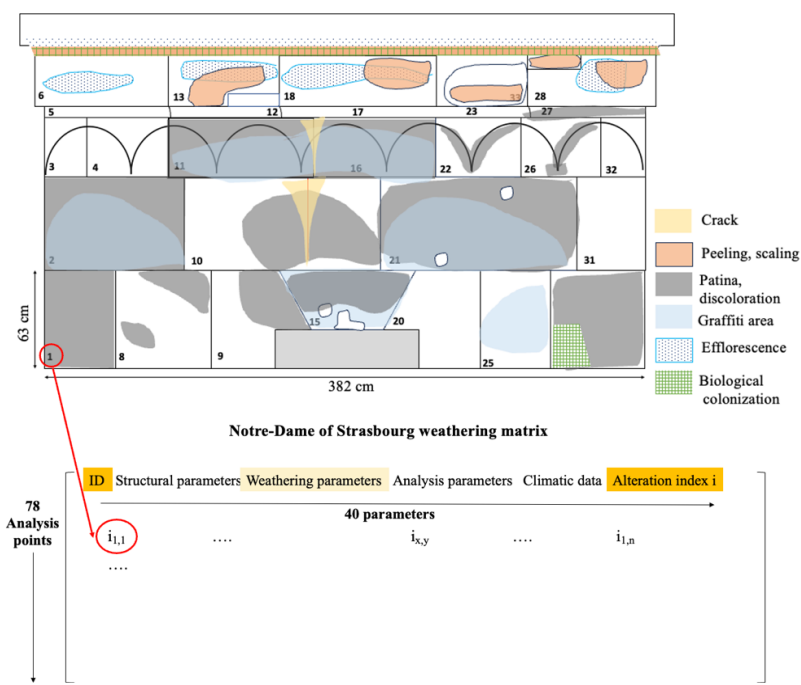

*Figure 5: Presentation of the general structure of the weathering matrix: 1 block > 1 ID > 1 line and weathering mapping of the Southwest face*

*Table 4: Rating and sub-categorisation of weathering forms for index calculation for each block (ID) of the matrix*

| Rating (i) | 0 | 1 | 2 | 3 | 4 | 5 |
|---|---|---|---|---|---|---|
| **Degree of affection** | Not altered | Covers 0-10% surface | Covers 10-25% surface | Covers 25-50% surface | Covers >75% surface | All area affected |
| **General aspect** | Not altered | Slightly altered | Moderately altered | Altered | Much altered | Severely altered |

## 4. Conclusion

To study the impact of climate change on the weathering of heritage materials, a climate monitoring is carried out on three cultural sites in France for three years. This paper presents the climatic instrumentation and the diagnosis of the initial state on the spire of Strasbourg cathedral. The first climatic data collected reveal a particularly extreme microclimate with significant daily variations in temperature and humidity. Whatever the face of the spire studied, the moisture in the sandstone seems to come from water flowing from above. This water may be loaded with solubilized salts that recrystallize at depth and on the surface during the soaking/drying cycles. Analyses of water and soluble salt content of powders at different depths identified the presence of efflorescence and a significant state of alteration. All collected data is gathered in an initial matrix by the mean of a calculated weathering index. This data processing method overcomes the problem of multimodality of data, with the use of supervised learning prediction models. Over time, monitoring the state of weathering will generate new matrices with an updated weathering index. New parameters will be added to complete the climatic instrumentation, such as wind speed or sunshine. Finally, this approach developed as part of the COLISEUM project could be applied to other sites, giving a tool for heritage conservation strategies in the context of climate change.

## Acknowledgment

As part of the ESPADON (En Sciences du Patrimoine, l'Analyse Dynamique des Objets anciens et Numériques) project (2021-2028), this work is supported by the French National Research Agency (ANR-21-ESRE-0050)- Fondation des sciences du patrimoine. It aims to enrich the developing interdisciplinary field of "Heritage Sciences". We thank David Roqui, and our co-supervisors from ETIS - CY Cergy Paris Université, C2RMF, CICRP, Mescla Patrimoine, for their help and contribution in the project. We thank the Fondation de l'Oeuvre Notre-Dame de Strasbourg, Bibracte site and the Prud'homie des pêcheurs of Saint-Pierre chapel for their help and for allowing us to instrument and study the sites.